\documentclass[twocolumn,showpacs,preprintnumbers,amsmath,amssymb]{revtex4}

\pdfoutput=1
\usepackage[latin1]{inputenc}
\usepackage[english]{babel}
\usepackage[T1]{fontenc}
\usepackage{graphicx,amsfonts}
\usepackage{amsmath}

\begin{document}

\title{Constant-net-time headway as key mechanism\\ behind pedestrian flow dynamics}

\author{Anders Johansson}
 \email{anders.johansson@gess.ethz.ch}
\affiliation{
ETH Zurich,\\
UNO C 11 Universit\"{a}tstrasse 41, 8092 Zurich, Switzerland\\
}

\date{\today}

\begin{abstract}
We show that keeping a constant lower limit on the net-time headway is the key mechanism
behind the dynamics of pedestrian streams.
There is a large variety in flow and speed as functions of density for empirical data of pedestrian streams,
obtained from studies in different countries. The net-time headway however, stays approximately constant over
all these different data sets.

By using this fact, we demonstrate how the underlying dynamics of pedestrian crowds,
naturally follows from local interactions.
This means that there is no need to come up with an arbitrary
fit function (with arbitrary fit parameters) as has traditionally been done.

Further, by using not only the average density values, but the
variance as well, we show how the recently reported stop-and-go
waves [Helbing et al., {\em Physical Review E}, {\bf 75}, 046109]
emerge when local density variations take values exceeding a certain
maximum global (average) density, which makes pedestrians stop.
\end{abstract}



\maketitle

\section{Introduction}

With an increasing population and with more cost effective transportation,
mass gatherings become more frequent. The total size of such gatherings
are often as large as millions of people, for example during the inauguration ceremony
of president Obama \cite{pop_mech_obama} and the Hajj pilgrimage
to Mecca \cite{HelbingEtAlMakkah}.

To guarantee the safety of the participants during such large mass
gatherings, careful planning needs to be carried out by the
organizer. During the last decades, numerous empirical studies
\cite{weidmann,Mori,Fruin,ando,Polus,smith_and_dickie,Still,Teknomo,SeyfriedFD,kretz,HelbingEtAlMakkah,ped_encyclopedia}
have been performed on pedestrian crowds in different countries, in
order to understand the dynamics of these crowds. Even though an
understanding of crowd dynamics is a prerequisite for being able to
plan a mass gathering, there is still no concensus on some of the
most basic relations, such as how the flow of people (people per
meter per second) depends on the crowd density (people per m$^2$).
Misconceptions of these basic relations may result in serious safety
risks during mass gatherings~\cite{ped_encyclopedia}.

Let us now start from the bottom up, and show how local interactions
lead to certain flow-density relationships for the stream of
pedestrians. Since movement and avoidance patterns of pedestrians
tend to be rather complex, the traditional way to reduce complexity
is to find a relation of the flow $Q$ ($m^{-1}s^{-1}$) as a function
of the average density $\varrho$ ($m^{-2}$). Using this relation,
called the {\em fundamental diagram},
has been successful to some extent, but unfortunately there are large variations of these
relations, among empirical studies carried out in different countries.
All these studies agree on that walking speed of pedestrians is a decreasing function of density,
but they disagree on how this function looks like. We will now demonstrate how the net-time headway,
as a result of finite reaction times, is the key mechanism which can explain the discrepancies
between data sets from different studies.

\section{Reaction time}

It is known from traffic science that finite reaction times are
needed to explain instabilities in traffic flows
\cite{kesting_delays}. For pedestrian-flow dynamics, the role of
finite reaction times has not been investigated in detail. By doing
so, it turns out that the finite reaction time gives rise to a
certain net-time headway, which is needed as a safety headway, to
prevent accidental physical encounters with surrounding pedestrians.

\begin{figure}
\begin{center}
    \includegraphics[width=0.5\textwidth]{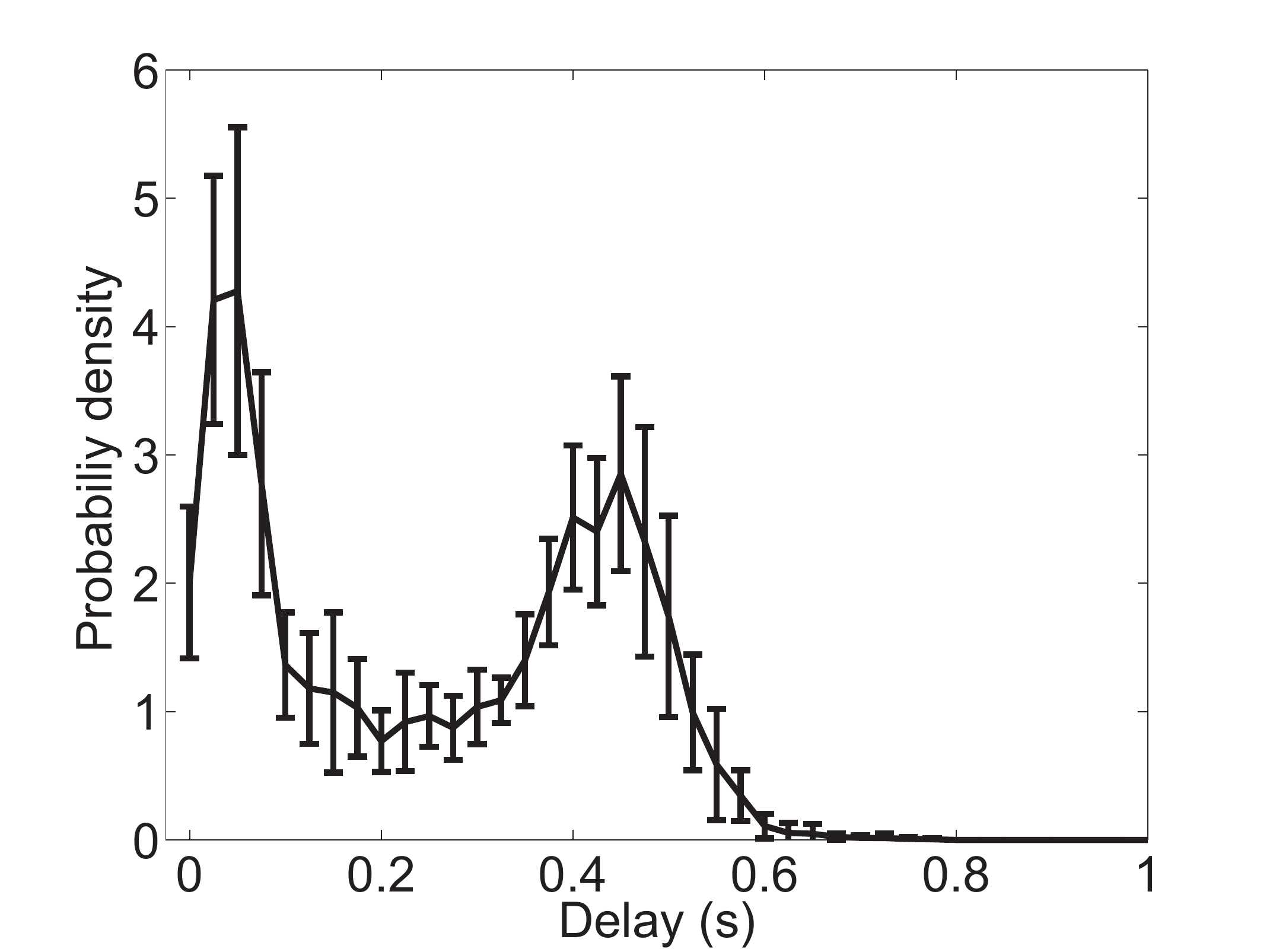}
\end{center}
\caption
{Probability-density function of the delay $T^d$.
The error bars correspond to one standard deviation.
Interestingly, the probability-density function is bi-modal.
When the surrounding pedestrians are acting in a way that is easy to predict,
extrapolation allows to anticipate their behaviors,
while a delayed reaction results in cases of unexpected behaviors.
}
\label{fig_delays}
\end{figure}

Many-particle simulations \cite{HelbMoln1995,panic} coupled with empirical pedestrian-trajectory data \cite{johansson_thesis}
reveal the probability-density function
of delay times $T^d$ from a walking experiment \cite{hoogendoorn_exp} where two pedestrian streams are intersecting
at a $90^\circ$ angle. The resulting distribution of delay times are shown in Fig.~\ref{fig_delays}.

Interestingly, the probability-density function of the delay curve
is bi-modal. The first peak occurs at lower times than the typical
response times, to visual or acoustic cues \cite{woodworth,welford}.
Therefore, this peak must correspond to \emph{anticipated} movements
of the surrounding pedestrians. The second peak at around 0.45 s
occurs at times which are significantly larger than the previously
mentioned response times, but also lower than response times
involving conscious reactions \cite{kesting_delays,green_delays}.
Therefore, we conclude that this second peak corresponds to an
unconscious response, which is more complex than a simple reaction.
In fact, it has been shown that reactions where there are more than
one possible response (choice reaction time) as well as reactions to
more complex cues (recognition reaction time) take significantly
longer time \cite{donders}.

We interpret the bi-modality as follows:
When the surrounding pedestrians act in a way that is easy to predict,
extrapolation allows to anticipate their behaviors,
while a delayed reaction results in cases of unexpected behaviors.

\section{Model}

There has been a rich
amount of microscopic models of pedestrian dynamics, for example the social-force model \cite{panic,HelbMoln1995}
and cellular-automata models \cite{ca,ca2,ca3}. These models are able to reproduce various self-organization phenomena,
such as lane and stripe formation \cite{transci}, freezing by heating \cite{freezingbyheating},
Mexican waves in excitable media \cite{mexicanwave},
intermittent outflows \cite{bottleneckPRL}, stop-and-go waves and crowd turbulence \cite{HelbMoln1995}.

When measuring empirical pedestrian flows and densities and then fitting a suitable curve to
the data, one obtains a function which is useful for engineering involved in planning of pedestrian facilities.
This pragmatic fit curve, however, does not provide any insight into the mechanisms and dynamics behind
the pedestrian interactions and behaviours, leading to the aggregated data.

\begin{figure}[htb] \begin{center}
    \includegraphics[width=0.435\textwidth]{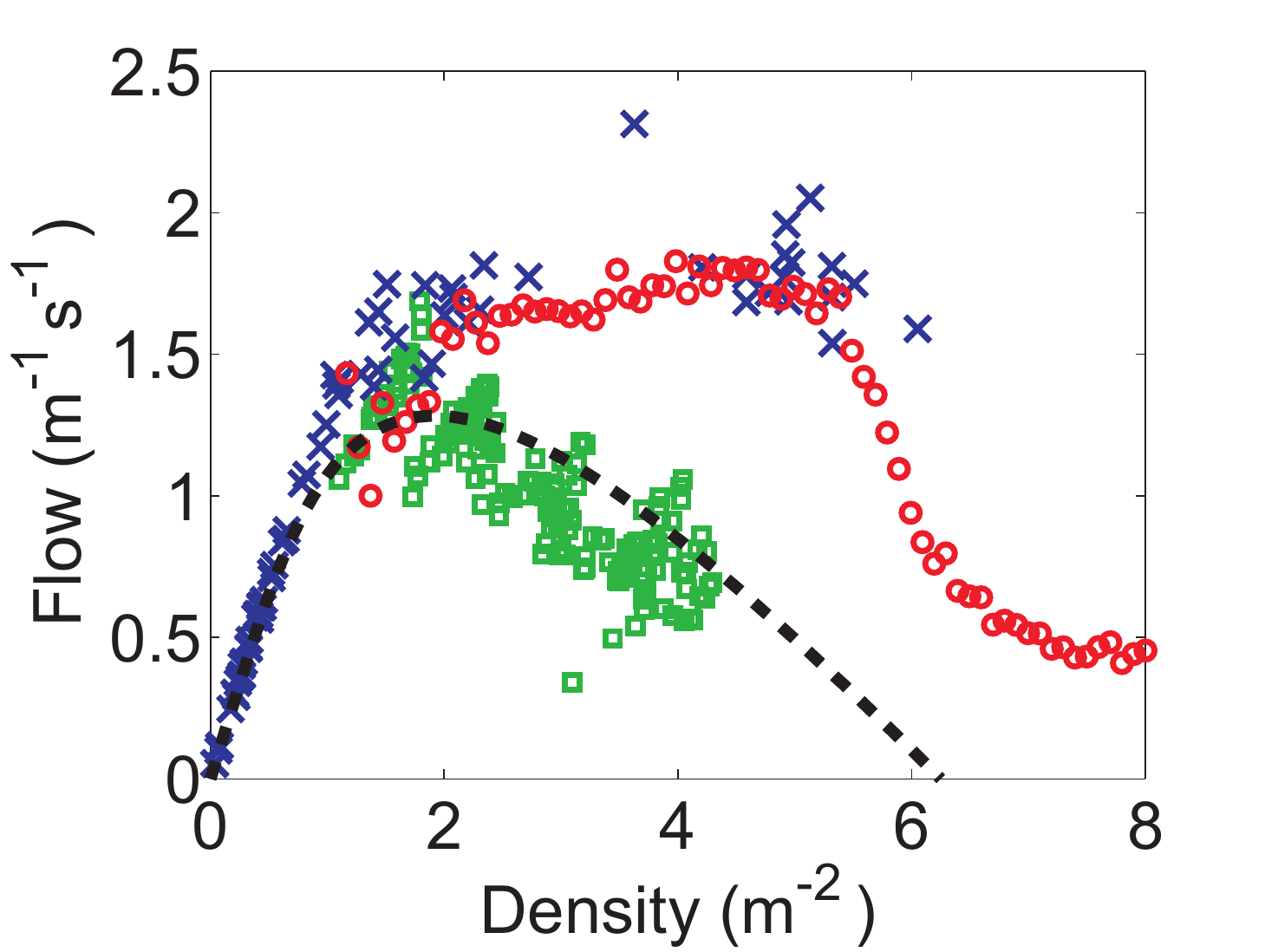}
    \includegraphics[width=0.435\textwidth]{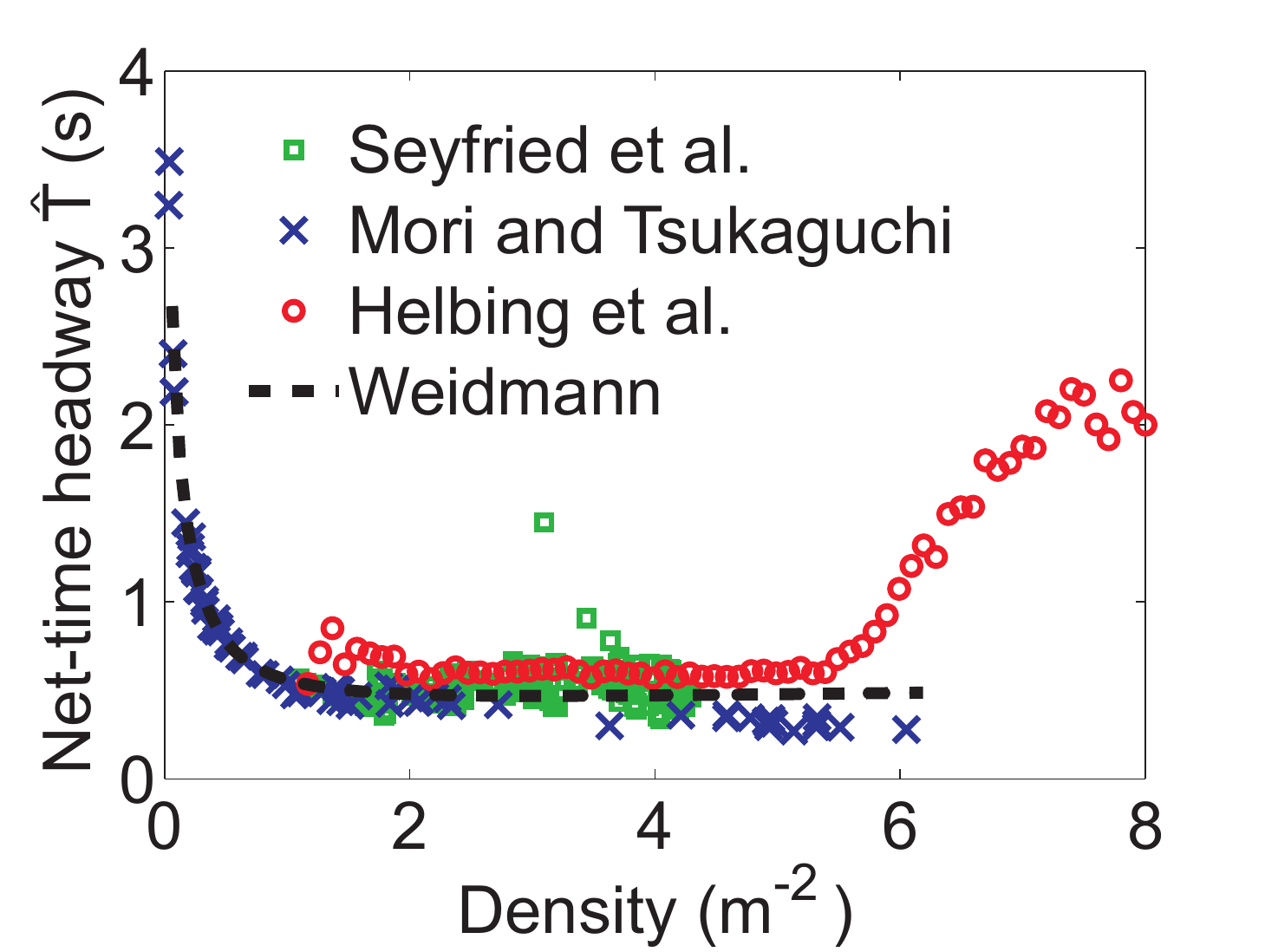}
    \caption
{(Color online) Top: Flow as a function of density, for data from a number of empirical studies.
    Bottom: The net-time headways $\hat T$ as a function of density $\varrho$.
    $\hat T$ is most often bounded by a constant lower value of about 0.5 s.
    In the data of Ref.~\cite{HelbingEtAlMakkah},
    however, there is a transition for high densities where $\hat T$ suddenly increases.
    The data sets are the same as used in Ref.~\cite{HelbingEtAlMakkah}, i.e. the data 
    from (Helbing et al.) correspond to {\em local} densities and flows.
    }
\label{fd_and_headways}
\end{center} \end{figure}

However, when plotting the fundamental diagrams obtained in various
empirical studies (Fig.~\ref{fd_and_headways} (top)), one can see
that each of the curves has a similar parabola-like shape.
Nevertheless, the curves are quite different from one measurement
site to another. One question remains to be answered: {\em What, if
any, is the common underlying principle of these curves?}

In an attempt to bridge this knowledge gap, let us come back to the
issue of reaction times, mentioned before. Since pedestrians have a
typical reaction time $T^d$=0.45 s to unexpected behaviors of
surrounding pedestrians, it would be natural that they compensate
the risk of bumping into others, by keeping a certain safety time
headway to the surrounding pedestrians 
\footnote{Note that we use the peak at longer delays since this peak
corresponds to {\em unexpected} events. Even though there is a peak
at lower delay times, there is no guarantee that the behaviours 
of others can always be anticipated, and therefore the
maximum peak must be used as a safety time.}.

\begin{figure}
\begin{center}
    \includegraphics[width=0.5\textwidth]{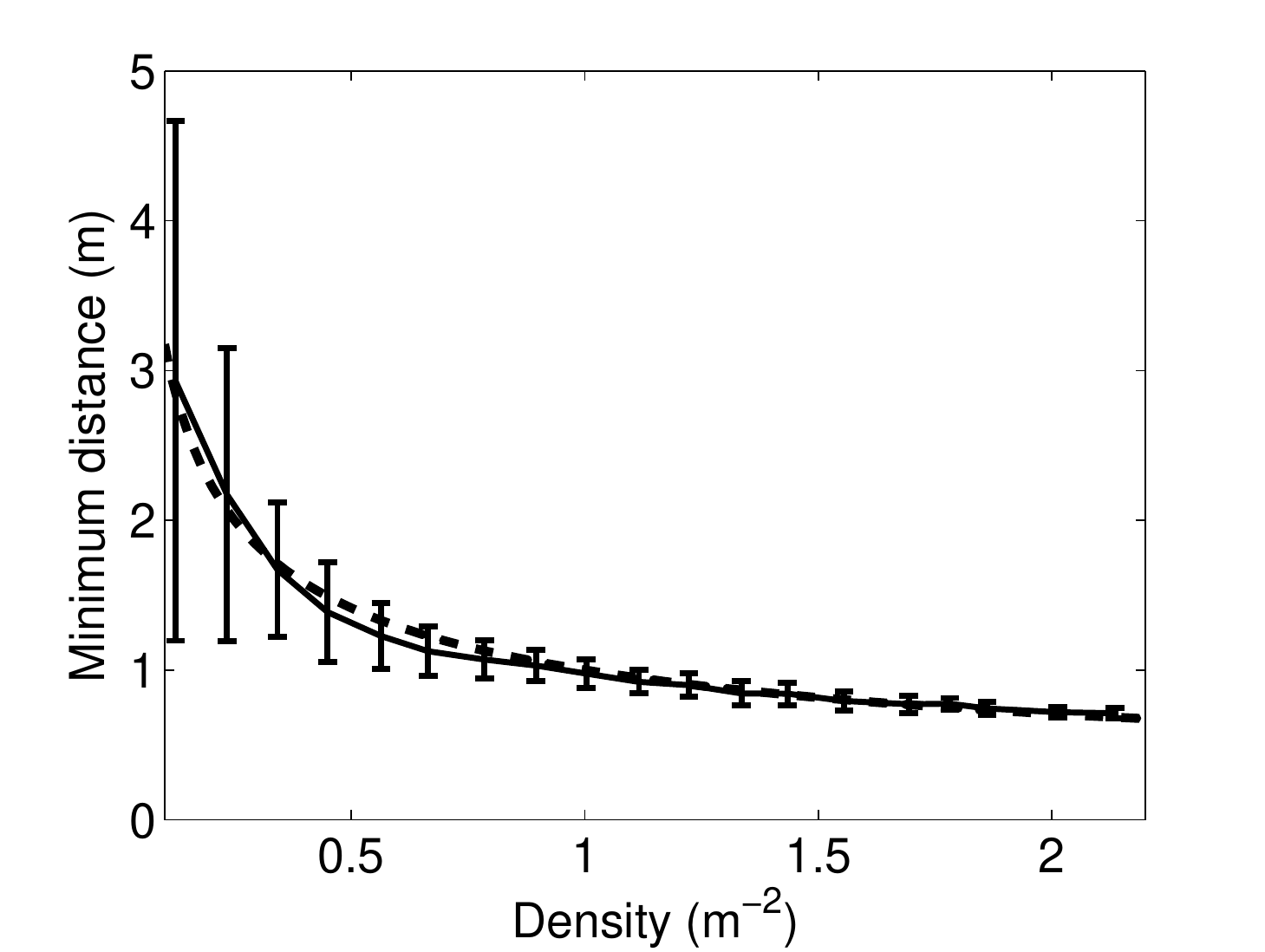}
\end{center}
\caption {The distance between an arbitrary pedestrian $\alpha$ and
the closest surrounding pedestrian $\beta$, as a function of global
(average) crowd density $\varrho$. The solid line shows the average
value $\pm$ one standard deviation as error bars, and the dashed
line shows the fit curve $1/\sqrt(\varrho)$. The data is from
Ref.~\cite{hoogendoorn_exp}} \label{fig_rho_vs_dist}
\end{figure}

To connect the aggregated density to local interactions, let us approximate the mean distance between the center-of-masses of a pedestrian $\alpha$
and the closest pedestrian $\beta$ by $d = \langle d_{\alpha \beta} \rangle = 1/\sqrt{\varrho}$,
where $\varrho$ is the global 
\footnote{The global density $\varrho$ is defined as the number of people within a certain area, divided by that area.
The local density $\rho$ \cite{HelbingEtAlMakkah} on the other hand, is defined via a bell-shaped weight function, where the influence
of close pedestrians is larger than the influence of remote pedestrians.}
(average) density. Note that this would hold only if the pedestrians were distributed into a square lattice, but for other density distributions, it will serve as a fair approximation (see Fig.~\ref{fig_rho_vs_dist}).

The net distance is defined as $\hat d = d - 2r$, where $r = 1/(2\sqrt{\varrho_{max})}$ is the effective radius of a pedestrian, and $\varrho_{max}$ is the largest measured density.

Assuming that the predecessor $\beta$ \footnote{This assumes a
following behaviour for uni-directional streams, which is most often
the case for empirical pedestrian studies.} would suddenly stop
\cite{bottleneckPRL}, it would take 
$\hat T={\hat d}/v_{\alpha} =(1/\sqrt{\varrho} - 1/\sqrt{\varrho_{max}})/v_{\alpha}$
seconds before a physical encounter with pedestrian $\alpha$ occurs,
if $v_{\alpha}$ is the speed of pedestrian $\alpha$. We now show how
the net-time headway $\hat T$ 
depends on the global density
$\varrho$ by applying the above scheme to empirical data determined
from different studies (see Fig.~\ref{fd_and_headways}(bottom)).

Note that $\hat T$ saturates at a constant value, that is very similar to the response time to unexpected behaviors (see Fig.~\ref{fig_delays}).
However, in the data of Ref.~\cite{HelbingEtAlMakkah} there is a transition at very high densities, where $\hat T$ suddenly \emph{increases}. This can be interpreted in at least two ways:
\begin{itemize}
    \item {\em Hypothesis 1}: When the density is very high, pedestrians start to have fear of crushing or asphyxia
    \cite{turbulencesim}, and therefore want to increase the space around themselves (leading to higher net-time headways $\hat T$).
    \item {\em Hypothesis 2}: If the space in front of a pedestrian is too small
    (or the velocity is too low) it will no longer be possible to take normal steps.
    Rather, pedestrians will completely stop until they have gained enough space to make a step.
\end{itemize}

In previous work \cite{turbulencesim}, {\em Hypothesis 1} has been used.
In this study, however, we will investigate {\em Hypothesis 2}.
This interpretation would naturally explain the empirically observed stop-and-go waves analyzed
in Ref.~\cite{HelbingEtAlMakkah}, and would further imply:
\emph{Above an average density of 5 persons per $m^2$, the fundamental diagram will no longer describe the
dynamics of the crowd well, since the flow rate is then alternating between movement
and standstill rather than continuous}.

In an attempt to unify all fundamental diagrams in the same framework, the following scheme is proposed: \\

Each pedestrian $\alpha$ has a free speed $v_{\alpha}^{0}=v_{max}$
(which is an upper speed limit, occurring when $\varrho \to 0$).
Each pedestrian also has a lower limit $v_{min}$ of the speed. For
$v<v_{min}$, pedestrians can no longer make normal steps, and would
rather stop completely. For simplicity, these values are assumed to
be the same for all pedestrians. It has been reported in
Ref.~\cite{hoogendoorn2005} that the free (unconstrained) headways
are exponentially distributed, where the constrained headways on the
other hand, are limited by a desired minimum headway. Therefore, we
propose that the net-time headway $\hat T$ is the key control
parameter for the fundamental diagram. That is, pedestrians will
decrease their speeds, if necessary, to assure a constant lower
limit of the net-time headway $\hat T$. The fundamental diagram can
now be specified as:
\begin{equation}
        v(\varrho) = \frac{d-2r}{\hat T} = \frac{1/\sqrt{\varrho}-1/\sqrt{\varrho_{max}}}{\hat T}
    \label{eq_v_vs_T}
\end{equation}
and bounded by $[v_{min}, v_{max}]$

\begin{figure}[htb] \begin{center}
    \includegraphics[width=0.5\textwidth]{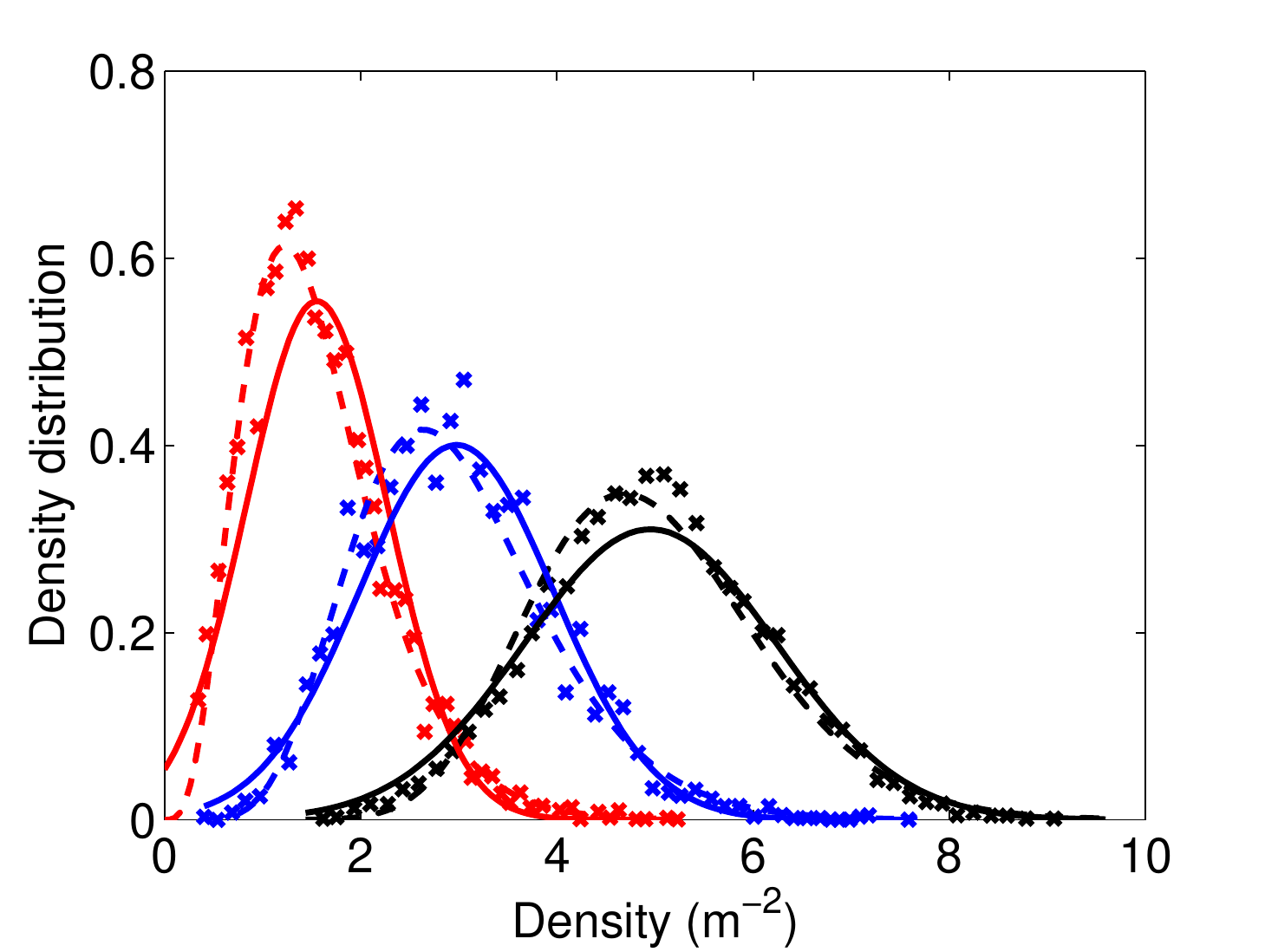}
    \caption{(Color online) 
    Distributions of local densities $\rho$ for three different global
 (average) densities $\varrho=1.6, 3$ and 5 pedestrians/$m^2$.
The data comes from Ref.~\cite{HelbingEtAlMakkah}. For each global density $\varrho$
a Beta distribution is fitted (dashed lines). However, Gaussian distributions (solid lines)
also fit the data fairly well. The Gaussian distributions are produced with the parameters
$\mu = \varrho$ and $\sigma=\sqrt{\varrho/3}$.
}
\label{fig_density_distr}
\end{center} \end{figure}

It has been shown in Ref.~\cite{johansson_acs_fundamental} that each
average density $\varrho$ corresponds to a distribution of local densities $\rho$,
and we therefore approximate the density distribution with a Gaussian distribution
$\mathcal{N}(\varrho,\sqrt{\varrho/3})$ with mean $\varrho$
and standard deviation $\sqrt{\varrho/3}$
(see Fig.~\ref{fig_density_distr}).

According to {\em Hypothesis 2}, defined above, we get an extra constraint,
saying that pedestrians will stop walking if they are too close to other pedestrians,
which happens for $\rho \ge \varrho_{max}$ (physical interaction).
They will then resume walking again when they have enough space $L$
for taking a step. Since one step (for low walking speed) needs  
approximately $L \approx  0.5$ m \cite{hoogendoorn2005}, we get a new 
net-time headway $\hat T'=L/v_{min} \approx 10$ s, whenever $\rho \ge
\varrho_{max}$.

\begin{figure}[!htbp]
\begin{center}
        \includegraphics[width=0.4\textwidth]{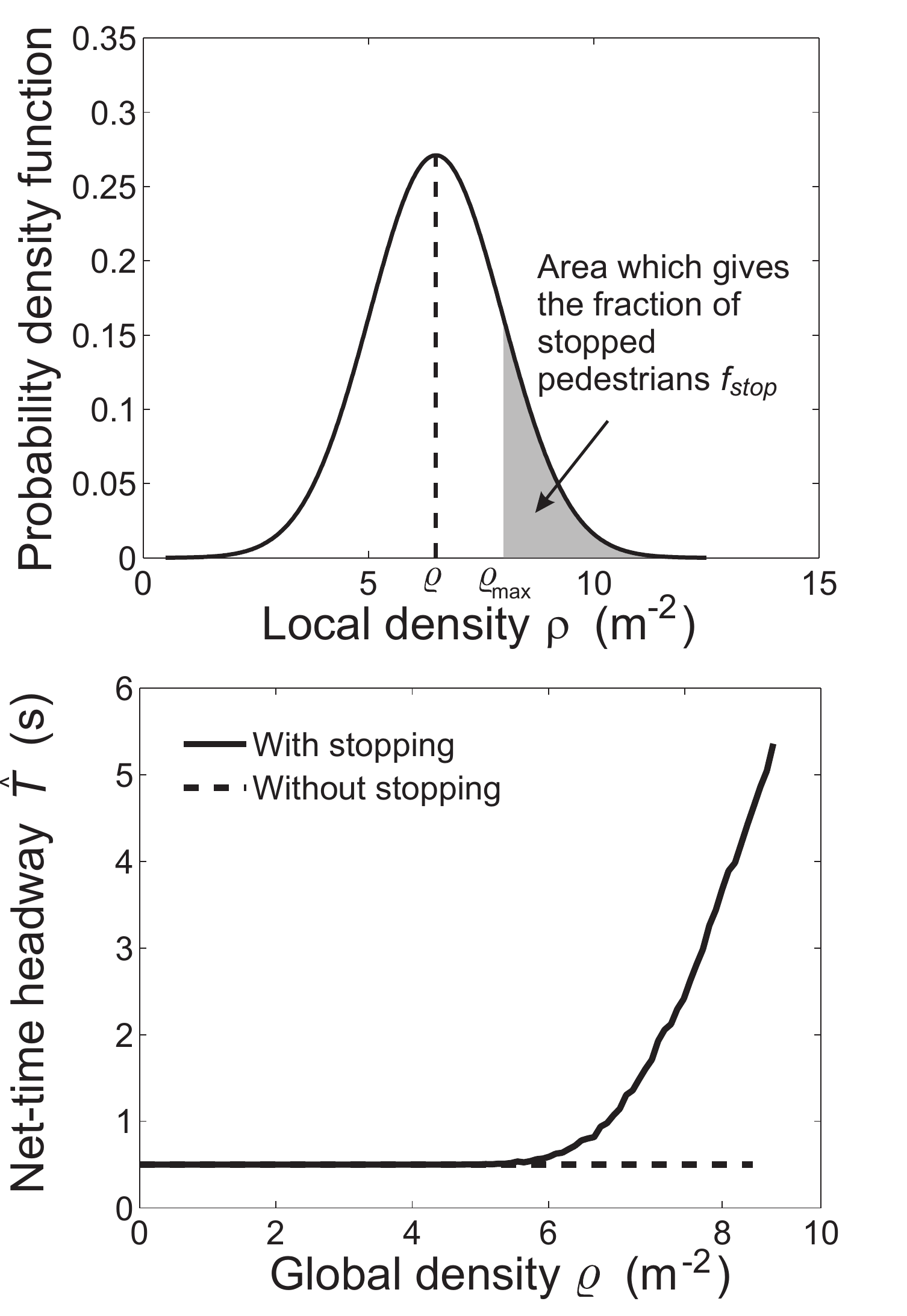}
    \caption{Top: The mean net-time headway $\langle \hat T \rangle$ is obtained
via the fraction of pedestrians who are physically colliding with
others, and is therefore stopping and temporarily increasing their
net-time headway. This fraction is obtained by integrating over the
probability-density-function of the local-density distribution,
starting at local densities $\rho$ that are higher than the maximum
global density $\varrho$. Bottom: The mean net-time-headway $\langle
\hat T \rangle$ (solid line) as a function of the global density
$\varrho$. The net-time headway without stopping is displayed as a
dashed line. } \label{area_and_net_time_headway}
\end{center}
\end{figure}

The fraction of pedestrians that are physically colliding with
others, can be measured by integrating the
probability-density-function of the Gaussian distribution (see
Fig.~\ref{area_and_net_time_headway} (top) ).
\begin{equation}
    f_{stop} = \int_{\varrho_{max}}^{\infty} \mathcal{N}(\rho) d \rho
\end{equation}
with mean $\mu=\varrho$ and standard deviation
$\sigma=\sqrt{\varrho/3}$. Then, the mean net-time headway (see
Fig.~\ref{area_and_net_time_headway} (bottom)) is given by the
fraction of stopped pedestrians as
\begin{equation}
    \langle \hat T \rangle = (1-f_{stop})\hat T + f_{stop} \frac{L}{v_{min}} \,.
    \label{eq_mean_T}
\end{equation}

\begin{figure}[!htbp]
\begin{center}
    \includegraphics[width=0.4\textwidth]{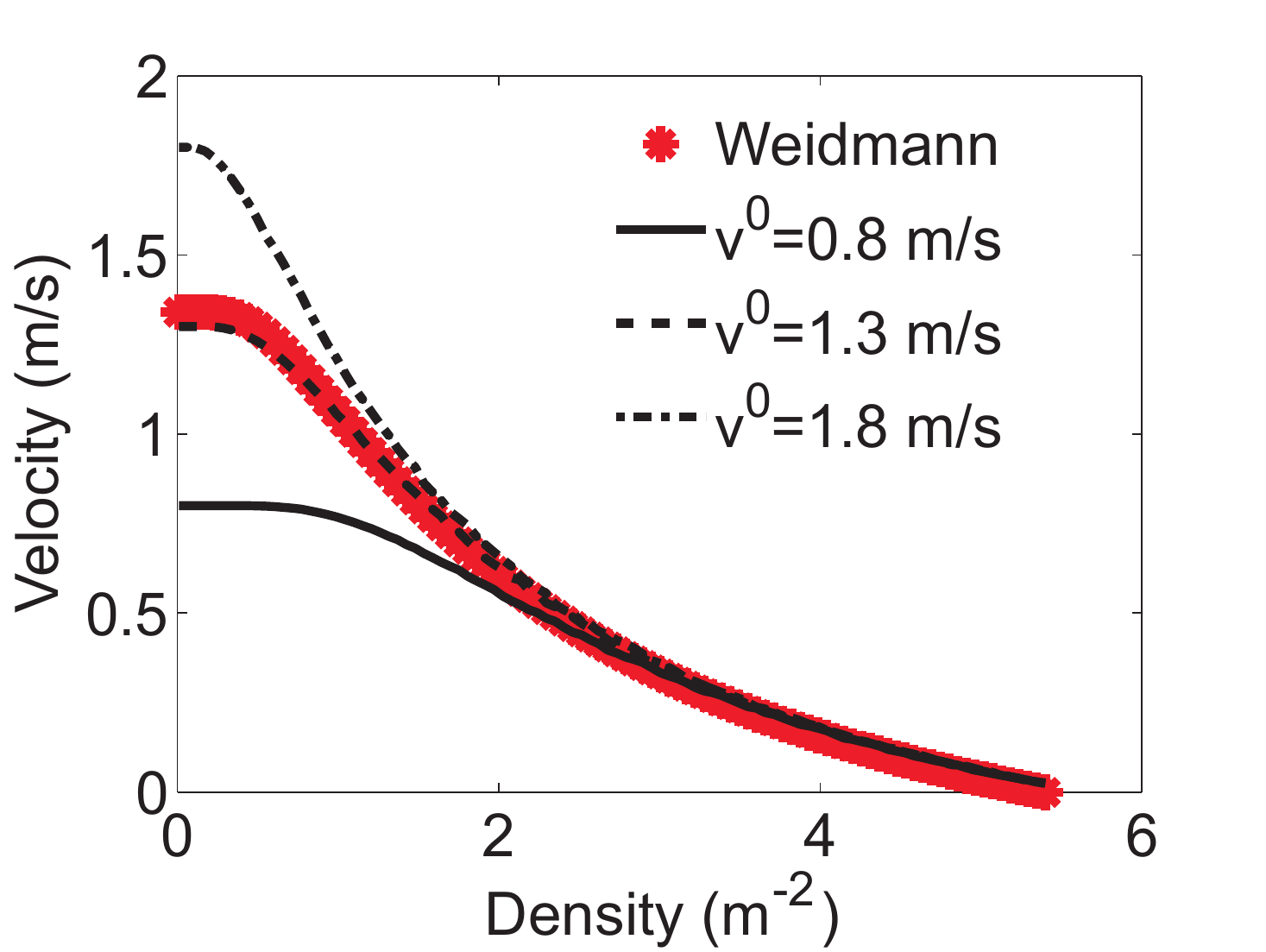}
    \includegraphics[width=0.4\textwidth]{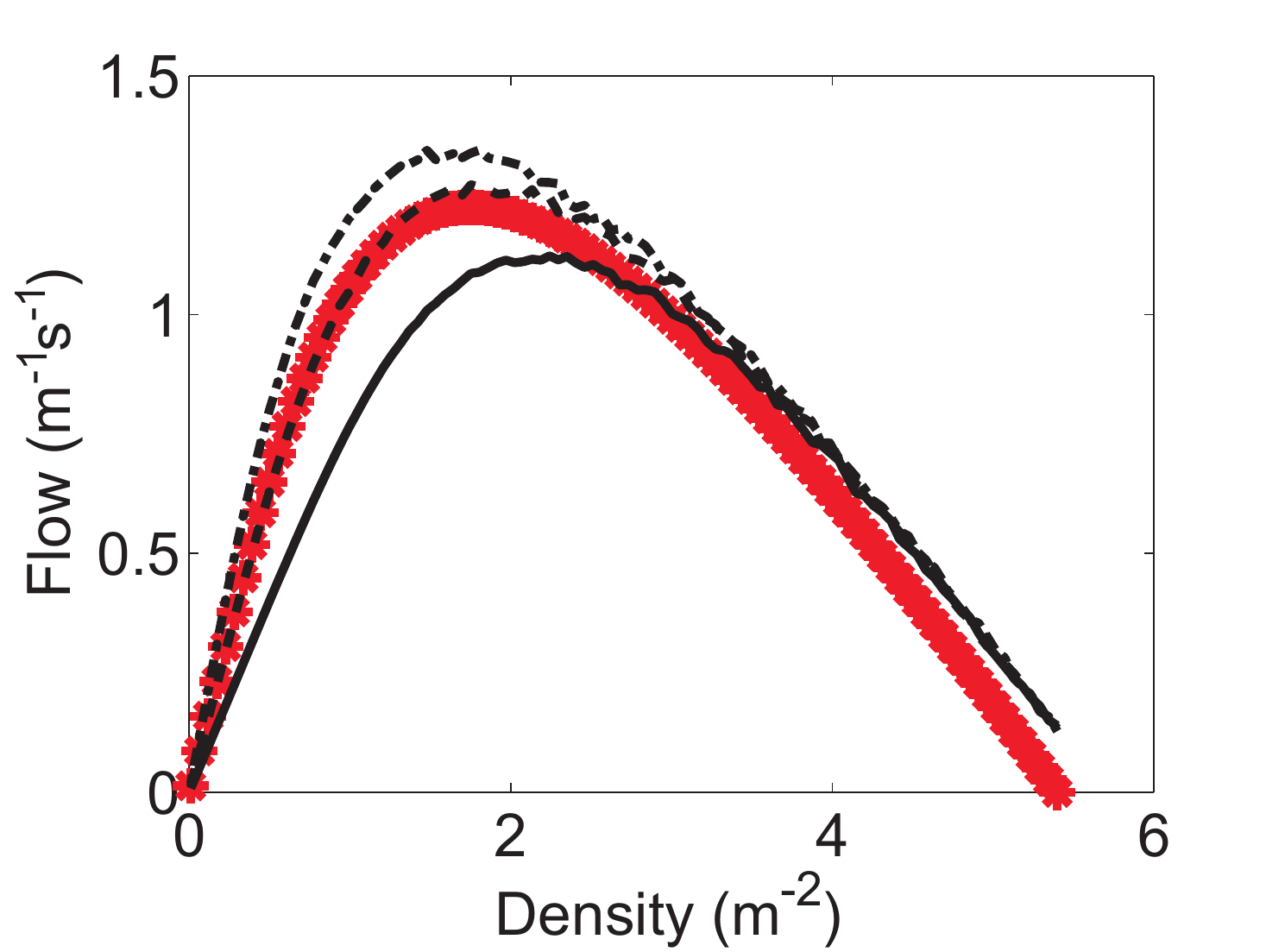}
    \caption{(Color online) Fundamental diagrams and velocity-density relations generated by
Eq.~(\ref{eq_v_vs_T}), for different free speeds $v^0$. Note how
they all converge for large densities. As a comparison, the
empirical fit curve by Weidmann \cite{weidmann} is shown. }
\label{fig_fd_v0}
\end{center}
\end{figure}

Figure~\ref{fig_fd_v0} shows
generated fundamental diagrams from Eq.~(\ref{eq_v_vs_T})
with the parameters $\hat T=0.5$ s, $\varrho_{max}=5.4$ m$^{-2}$,
and for different values of the free speed $v^0=v_{max}$. Since $v^0$ only gives
the upper limit of the velocity, fundamental diagrams with different $v^0$ converge
at high enough crowd densities, given that all other parameters are fixed. 
The reason is that, for high density, the movement is transformed from 
individual walking to walking which is constrained by other pedestrians.

\par

\par

\par\begin{figure}[!htbp]
\begin{center}
    \includegraphics[width=0.49\textwidth]{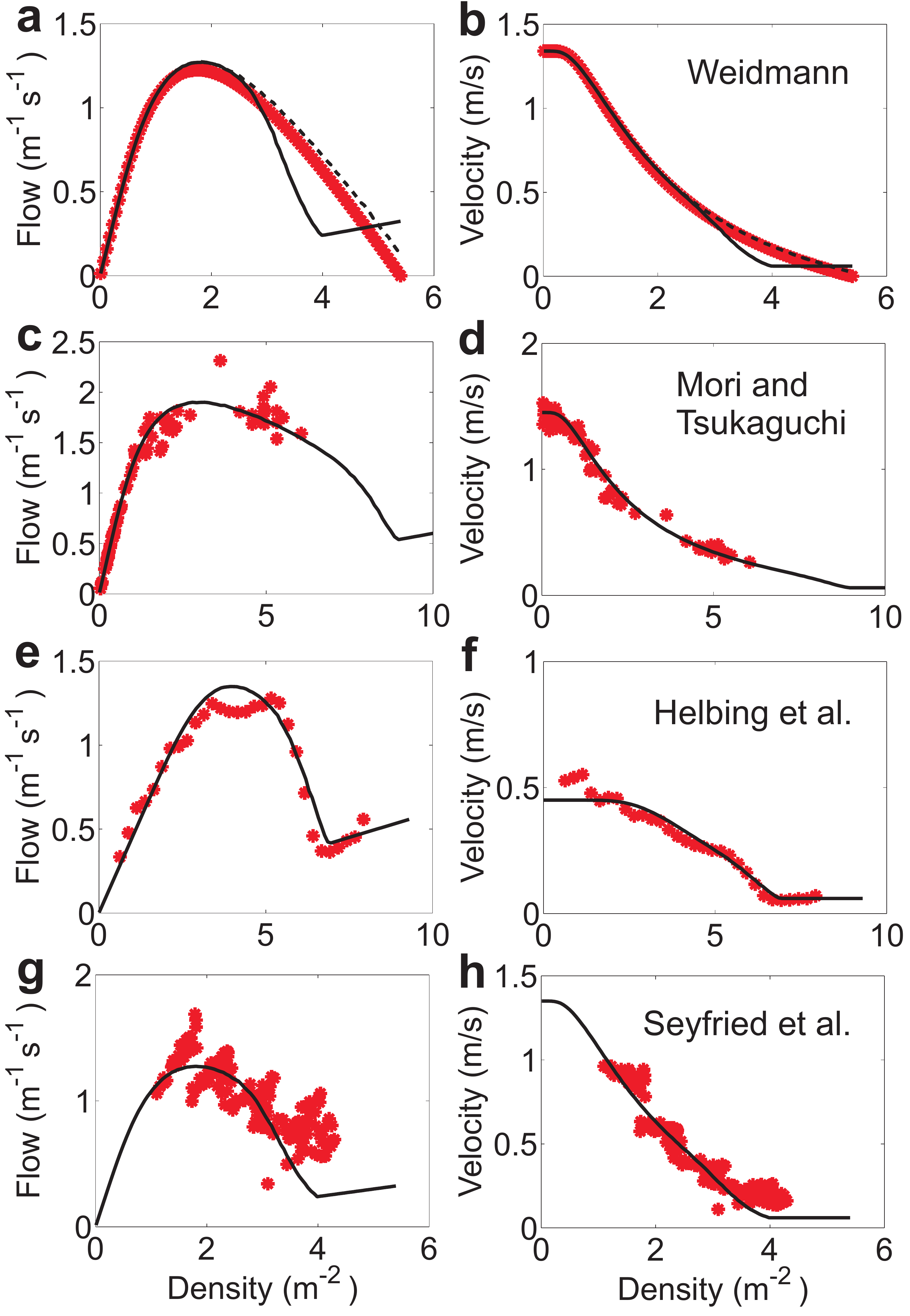}
    \caption{(Color online) 
Fundamental diagrams and velocity-density relations generated by Eqs.~(\ref{eq_v_vs_T}) to (\ref{eq_mean_T}),
assuming a constant net-time headway $\hat T=0.5$ s
and a minimum velocity $v_{min}=0.06$ m/s.
Red markers represent empirical data and solid lines the theoretically expected relationships.
a) Flow-density data, and b) velocity-density data by Weidmann \cite{weidmann}, compared to
a fundamental diagram generated with the parameters
$\varrho_{max}=5.4$ m$^{-2}$ and $v_{max}=1.34$ m/s. The dashed line shows the result when it is
assumed that no pedestrians are stopping, i.e. $f_{stop}=0$.
c) Flow-density data, and d) velocity-density data by Mori and Tsukaguchi \cite{Mori},
compared to a fundamental diagram generated with the parameters
$\varrho_{max}=12$ m$^{-2}$ and $v_{max}=1.45$ m/s.
e) Flow-density data, and f) velocity-density data from Helbing {\em et al.}~\cite{HelbingEtAlMakkah}, compared to a
fundamental diagram generated with the parameters $\varrho_{max}=9.3$ m$^{-2}$ and $v_{max}=0.45$ m/s.
g) Flow-density data, and h) velocity-density data from Seyfried {\em et al.}~\cite{SeyfriedFD}, compared to a
fundamental diagram generated with the parameters $\varrho_{max}=5.4$ m$^{-2}$ and $v_{max}=1.35$ m/s.
}
\label{fd_fits}
\end{center}
\end{figure}

We now apply the method outlined above on different empirical fundamental diagrams.
In all cases we use $\hat T = 0.5$ s and $v_{min}=0.06$ m/s.
Starting with Weidmann's \cite{weidmann} fundamental diagram, we have the parameters
$\varrho_{max}=5.4$ m$^{-2}$ and $v^{0}=1.34$ m/s, which is displayed in Fig.~\ref{fd_fits} (a, b)
together with our curve, obtained by Eqs.~(\ref{eq_v_vs_T}) to (\ref{eq_mean_T}).

Next, we apply our method to the fundamental diagrams of
Refs.~\cite{Mori}, \cite{HelbingEtAlMakkah} and \cite{SeyfriedFD}, and obtain the
results presented in Fig.~\ref{fd_fits} (c-h). The fit
functions match all four different empirical data sets well. All
parameters are kept constant over the different data sets, except
the maximum density and the free speed, but these two values are
obtained from the data, rather than tuned in order to fit the data.

\section{Conclusions}

The constant net-time headway is a natural safety mechanism to
compensate for the reaction time to unexpected events. It has been
demonstrated that various data sets, from different countries, all
share the same net-time headway $\hat T$ = 0.5 s. The particular
advantage of our method is that it follows naturally, without the
need of an arbitrary fit function. Further, all the parameters are
measurables, such as the free speed and the maximum density. There
is not a single free parameter that must be tuned in order to fit
the different data sets. However, it should be mentioned that even 
though the maximum density can be estimated, it can normally not 
be exactly determined from the data. This is addressed in 
recent work ~\cite{chattaraj} that may make it possible to obtain 
culturally dependent parameters, such as the maximum density.

\section{Acknowledgments}
The author is grateful for the partial financial support by the DFG grant He
2789/7-1.

He would also like to thank Dirk Helbing for his valuable comments,
Serge Hoogendoorn for sharing data from his experiments, and to Habib Al-Abideen
for sharing data from the Hajj.


\begin{thebibliography}{99}

\bibitem{pop_mech_obama}
P. Taylor,
How Officials Will Control the Crowds at Obama's Inauguration,
Popular Mechanics, January 20, 2009.

\bibitem{HelbingEtAlMakkah}
D. Helbing, A. Johansson, and H.~Z, Al-Abideen,
The Dynamics of Crowd Disasters: An Empirical Study,
Phys. Rev. E {\bf 75}, 046109 (2007).

\bibitem{weidmann}
U. Weidmann,
Transporttechnik der Fu{\ss}g\"anger,
{\em ETH-Z\"urich, Schriftenreihe IVT-Berichte} 90 (1993).

\bibitem{Mori}
M. Mori H. Tsukaguchi,
A new method for evaluation of level of service in pedestrian facilities,
Transportation Research A {\bf 21}(3), pp.~223--234 (1987).

\bibitem{Fruin}
J. J. Fruin,
Designing for pedestrians: A level-of-service concept,
Highway Research Record {\bf 355}, 1-15 (1971).

\bibitem{ando}
K. Ando, H. Ota, and T. Oki,
Forecasting the flow of people,
Railway Research Review {\bf 45}, 8-14 (1988).
(in Japanese).

\bibitem{Polus}
A. Polus, J. L. Schofer, and A. Ushpiz,
Pedestrian flow and level of service,
Journal of Transportation Engineering {\bf 109}, 46-56 (1983).

\bibitem{smith_and_dickie}
R. A. Smith and J. F. Dickie (eds.),
Engineering for Crowd Safety. (Elsevier, Amsterdam) (1993).

\bibitem{Still}
K. Still,
Crowd Dynamics,
PhD Thesis (2000).

\bibitem{Teknomo}
K. Teknomo,
Microscopic Pedestrian Flow Characteristics: Development of an Image Processing Data Collection and Simulation Model.
PhD Thesis, Japan (2002).

\bibitem{SeyfriedFD}
A. Seyfried, B. Steffen, W. Klingsch, and M. Boltes,
The fundamental diagram of pedestrian movement revisited,
J. Stat. Mech. P10002 (2005).

\bibitem{kretz}
T. Kretz, A. Gr\"unebohm, and M. Schreckenberg,
Experimental study of pedestrian flow through a bottleneck,
J. Stat. Mech P10014 (2006).

\bibitem{ped_encyclopedia}
A. Schadschneider, W. Klingsch, H. Kluepfel, T. Kretz, C. Rogsch, and A. Seyfried,
Evacuation Dynamics: Empirical Results, Modeling and Applications,
{\em Encyclopedia of Complexity and System Science, B. Meyers (Ed.)} (Springer, Berlin, 2008).

\bibitem{kesting_delays}
M. Treiber, A. Kesting, and D. Helbing,
Delays, inaccuracies and anticipation in microscopic traffic models,
Physica A {\bf 360}, pp.~71--88 (2006).

\bibitem{HelbMoln1995}
D. Helbing and P. Moln\'{a}r,
Social force model for pedestrian dynamics,
Phys, Rev. E {\bf 51}, pp.~4282-4286 (1995).

\bibitem{panic}
D. Helbing, I. Farkas, and T. Vicsek,
Simulating dynamical features of escape panic,
Nature {\bf 407}, pp.~487--490 (2000).

\bibitem{johansson_thesis}
A. Johansson, 
Data-driven modeling of pedestrian crowds,
PhD Thesis, TU Dresden (2009).

\bibitem{hoogendoorn_exp}
S.~P. Hoogendoorn, W. Daamen, and P.~H.~L. Bovy,
Extracting microscopic pedestrian characteristics from video data,
{\em Annual Meeting Transportation Res. Board Pre-print CD-Rom},
(Mira Digital Publishing, Washington, D.C) (2003).

\bibitem{woodworth}
R.~S. Woodworth and H. Schlosberg,
Experimental Psychology. Henry Holt, New York (1954).

\bibitem{welford}
A.~T. Welford,
Choice reaction time: Basic concepts.
In A. T. Welford (Ed.), Reaction Times. Academic Press, New York, pp.~73--128 (1980).

\bibitem{green_delays}
M. Green,
How Long Does It Take to Stop? Methodological analysis of driver perception-brake times,
Transport. Hum. Factors 2 pp.~195--216, (2000).

\bibitem{donders}
F.~C. Donders,
On the speed of mental processes.
Translated by W. G. Koster, 1969. Acta Psychologica, {\bf 30}, pp.~412--431 (1868).

\bibitem{ca}
A. Kirchner, K. Nishinari, and A. Schadschneider,
Friction effects and clogging in a cellular automaton model for pedestrian dynamics,
Phys. Rev. E {\bf 67}, 056122 (2003).

\bibitem{ca2}
C. Burstedde, K. Klauck, A. Schadschneider, and J. Zittartz,
Simulation of pedestrian dynamics using a two-dimensional cellular automaton,
Physica A, {\bf 295}(3-4), pp.~507--525 (2001).

\bibitem{ca3}
W. G. Weng, T. Chen, H. Y. Yuan, and W. C. Fan,
Cellular automaton simulation of pedestrian counter flow with different walk velocities,
Phys. Rev. E {\bf 74}, 036102 (2006)

\bibitem{transci}
D. Helbing, L. Buzna, A. Johansson, T. Werner,
Self-organized pedestrian crowd dynamics: Experiments, simulations, and design solutions,
Transp. Sci. {\bf 39}, pp.~1--24 (2005).

\bibitem{freezingbyheating}
D. Helbing, I. Farkas, and T. Vicsek,
Freezing by heating in a driven mesoscopic system,
Phys. Rev. Lett, {\bf 84}, pp.~1240--1243 (2000).

\bibitem{mexicanwave}
I. Farkas, D. Helbing, and T. Vicsek,
Mexican waves in an excitable medium,
Nature {\bf 419}, pp.~131--132 (2002).

\bibitem{bottleneckPRL}
D. Helbing, A. Johansson, J. Mathiesen, H.~M. Jensen, and A. Hansen,
Analytical approach to continuous and intermittent bottleneck flows,
Phys. Rev. Lett. {\bf 97}, 168001 (2006).

\bibitem{turbulencesim}
W. Yu and A. Johansson,
Modeling crowd turbulence by many-particle simulations,
Phys. Rev. E {\bf 76}, 046105 (2007).

\bibitem{hoogendoorn2005}
S.~P. Hoogendoorn, W. Daamen,
Pedestrian behavior at bottlenecks,
Transp. Sci. {\bf 39}(2), pp.~147--159 (2005).

\bibitem{johansson_acs_fundamental}
A. Johansson, D. Helbing, H.~Z. A-Abideen, S. Al-Bosta,
From crowd dynamics to crowd safety: A video-based analysis,
Advances in Complex Systems {\bf 11}(4), pp.~497--527 (2008).

\bibitem{chattaraj}
U. Chattaraj, A. Seyfried, and P. Chakroborty,
Comparison of pedestrian fundamental diagram across cultures,
Advances in Complex Systems {\bf 12}, 393 (2009).

\end{thebibliography}
\end{document}